# Volume nanograting formation in laser-silica interaction as a result of the 1D plasma-resonance ionization instability

V. B. Gildenburg[1,2,*], I. A. Pavlichenko[1,2]

[1]*University of Nizhny Novgorod, Nizhny Novgorod 603950, Russia*
[2]*Institute of Applied Physics, Russian Academy of Sciences, Nizhny Novgorod 603950, Russia*

The initial stage of the small-scale ionization-induced instability developing inside the fused silica volume exposed to the femtosecond laser pulse is studied as a possible initial cause of the self-organized nanograting formation. We have calculated the spatial spectra of the instability with the electron-hole diffusion taken into account for the first time and have found that it results in the formation of some hybrid (diffusion–wave) 1D structure with the spatial period determined as geometrical mean of the laser wavelength and characteristic diffusion length of the process considered. Near the threshold of the instability this period occurs to be approximately equal to the laser half-wavelength in the silica, close to the one experimentally observed.

Formation of periodic nanostructures ("nanogratings") inside the volume of the transparent dielectric by the series of the femtosecond laser pulses of the moderate intensity has attracted considerable attention in the last two decades as a perspective method for the high density optical information writing and storage [1-7]. The carried out experimental and theoretical studies have allowed to reveal some important features, dependences, and application conditions of this phenomenon (mainly for the structures formed in the fused silica). Nevertheless, the physical mechanisms responsible for the creation of the observed self-organized bulk nanostructures is still under discussion [6,7]. In particular, up to now, there are no theoretical models that would allow to explain clearly the high enough degree of regularity of this structure and calculate its spatial period (measured experimentally as some part of the optical wavelength $\lambda$) as function of external conditions and parameters. Meanwhile, the main mechanisms governing the periodicity of the ionization structure in the every unit laser pulse (that determines ultimately the periodicity in the induced nanograting) can likely be understood and described in a consistent and non-contradictory manner within the framework of known approach developed in studies of small-scale (subwave) ionization–field instabilities of electromagnetic wave in gases [8-12]. In the model we propose below, the instability arises at the stage of ionization (or its saturation at the undercritical plasma density level) due to mutual enhancement of small (seeding) 1D periodic perturbations of the electric field amplitude and the free electron density when the perturbation gradient is parallel to the electric field of the undisturbed plane wave producing the ionization.

Some other (3D) local field enhancement effects were considered in connection with the grating formation problem in the Refs. [2,3,13,14] for the case when the optical breakdown in silica develops around multiple seats of primary ionization (randomly localized small regions with strongly enhanced ionization rate). The numerical calculations fulfilled in [13,14] show that the interaction and competition of the arising 3D inhomogeneities leads in this case to the formation of enough complicated density-modulated structure. Though the tendency to the spatial modulation in the direction of the laser electric field with the period of the order of the laser wavelength is really traced in the resulted structure, it nevertheless differs considerably from the 1D regular periodic grating and contains additional (rather severe and irregular) spatial modulation in the directions perpendicular to the electric field. Probably, the given structure can be considered as that one formed at the nonlinear stage of the above-mentioned ionization–field instability under conditions when the only nonlocal (spatial-scale-dependent) mechanism governing the instability is the electromagnetic waves interference.

The second important nonlocality mechanism, namely, diffusion of the free charge carriers, was not taken into account, what excluded the possibility of the spatial spectrum restriction on the small scales side at the initial (linear) stage of the instability. As a result, the instability went to its nonlinear stage with strong (3D) perturbations in the broad area of its spatial spectrum predetermining the complicity and irregularity of the structure obtained.

Unlike the model considered in Ref. [13,14] we take into account in our model both above mechanisms of nonlocality. Both of them tend to depress the instability growth rate but have opposite characters in the spatial scale dependences, resulting, as we shall see, in the existence of the instability threshold and some optimal spatial period (corresponding to the maximum of the growth rate), thus promoting the formation of the 1D quasi-periodical structure at the very early (linear) stage of the instability.

For a sufficiently broad class of the ionizable condensed media, the spatio-temporal evolution of the field and plasma in optical discharges created by the laser pulse of moderate intensity can be described by the vector wave equation

$$\frac{2i\omega}{c^2}\frac{\partial \mathbf{E}}{\partial t} + \Delta \mathbf{E} + \nabla\left(\frac{\mathbf{E}}{\varepsilon}\nabla\varepsilon\right) + k^2 \varepsilon \mathbf{E} = 0 \qquad (1)$$

for the slow time-varying envelope of the electric field (represented as $\text{Re}[\mathbf{E}(\mathbf{r},t)\exp(-i\omega t)]$) and the continuity equation

$$\frac{\partial N}{\partial t} = D\,\Delta N + W(|\mathbf{E}|, N) \qquad (2)$$

for the free charge carriers (electron-hole plasma) density $N(\mathbf{r},t)$. In these equations $k = \omega/c$ is the laser wave number in vacuum; $\varepsilon = \varepsilon_m - (N/N_c)/(1+i\nu/\omega)$ is the dielectric function of the ionized medium; $\varepsilon_m$ is its value in the absence of ionization; $N_c = m\omega^2/(4\pi e^2)$ is the critical density of the produced plasma, $\omega$ is the field frequency; $\nu$ is the effective collision frequency (the inverse time of the carriers momentum relaxation); $D$ is the diffusion coefficient; the function $W(|\mathbf{E}|, N)$ is an algebraic sum of local rates of ionization and recombination. Different types of large scale discharge structures (smooth in the wavelength scale) were calculated based on the similar equation system in Refs. [6,15,16] (without the electron diffusion and small-scale initial modulation of the plasma density taken into account). However, these equations allow to describe the small-scale instability of interest as well.

Let us consider the evolution of the small-scale perturbations of the field and plasma against the background of some smooth (quasi-homogeneous and quasi-steady) discharge state created by the laser pulse. The spatial and temporal scales of this background state are supposed to be large enough as compared to those ones for the small perturbations of interest, so that the optical field $\mathbf{E}_0$ in this state can be treated as an $x$-polarized locally plane wave of a given amplitude $E_0$ propagating along the $z$ axis in a homogeneous ionized medium with a given plasma density $N_0 < N_c$ (under assuming the full compensation of the absorption and convergence/divergence processes):

$$\mathbf{E}_0 = \mathbf{x}_0 E_0 \exp(ik\sqrt{\varepsilon_0}z), \quad \varepsilon_0 = \varepsilon_m - (N/N_c). \qquad (3)$$

Next, by setting $N = N_0 + N_1$ and $E_x = (E_0 + E_1)\exp(ikz)$ for the perturbed state and performing simple algebra within the framework of the standard procedure of the stability analysis for small spatially-periodic perturbations of the form $\{N_1, E_1\} \propto \exp(\gamma t - i\mathbf{q}\cdot\mathbf{r})$, we obtain the following characteristic equation relating the time constant (growth rate) $\gamma$ of the perturbation and its wave vector $\mathbf{q} = \mathbf{x}_0 q_x + \mathbf{y}_0 q_y + \mathbf{z}_0 q_z$ (see also Ref. [8]):

$$(\gamma + \alpha + Dq^2)\left[\frac{q^2}{k^2} + 4\frac{k^2}{q^2}\left(\frac{\gamma}{\omega} + i\frac{q_z}{k}\sqrt{\varepsilon_0}\right)^2\right] = \beta\left[\frac{q_x^2}{k^2}\mathrm{Re}\left(\frac{1}{\varepsilon}\right) - 1\right]. \qquad (4)$$

Here

$$\alpha = -(dW/dN)_0, \quad \beta = (1/N_c)E_0(dW/d|\mathbf{E}|)_0, \qquad (5)$$

the derivatives of the full ionization rate $W(|\mathbf{E}|, N)$ are taken for the undisturbed state.

Among the various types of instabilities corresponding to the different roots of this characteristic equation, the object of our main interest here is so called aperiodical "plasma-resonance-induced instability" (PRI) [8,9,11], in which the density perturbations increases or decreases monotonically in time at every space point throughout the laser pulse duration. It is precisely this case where the density modulation is pure transverse and the immobile 1D plasma grating arises ($q_z = 0$, $\text{Im}\,\gamma = 0$). Under inequalities $\gamma/\omega \ll (q/k)^2$, $\varepsilon_0 \gg \nu/\omega$, that we assume to be fulfilled in what follows, the mutual enhancement action between small perturbations of the electric field and the plasma density and corresponding root of the characteristic equation, as it is follows from Eqs. (1) - (4), are described for this instability by the relations ($q^2 = q_x^2 + q_y^2$)

$$E_1 = \frac{q_x^2 - k^2 \varepsilon_0}{\varepsilon_0 q^2} N_1, \quad (6)$$

$$N_1 = \frac{\beta}{\gamma + q_x^2 D + \alpha} E_1, \quad (7)$$

$$\gamma = -(\alpha + Dq^2) + \beta \frac{q_x^2 - k^2 \varepsilon_0}{q^2 \varepsilon_0} \quad (8)$$

Unlike the instabilities caused by the processes of ionization-induced scattering [8,10,12] which correspond to other roots of Eq. (4), the instability of interest has a quasi-static nature. It is associated with no wave synchronism conditions and is not damped by the convection of the wave perturbations from the ionization region. The positive feed-back loop mechanism for it is connected with the plasma resonance effect and can be understood with a simple "plasma-slab-condencer" model: as long as the plasma density remains lower the critical one, its small growth within a thin (subwave-length) plain layer leads to the local increase in the normal component of the electric field amplitude ($E_x \sim \text{const}/\varepsilon$) and therefore to the growth of the ionization rate and to the further growth of the density. Though the above mentioned "ionization scattering instability" is characterized by the higher typical values of the growth rate ($\gamma \sim \sqrt{i\omega W/N_c}$ for the backward scattering with $q_z \approx -2k\sqrt{\varepsilon_0}$ [8]), its manifestation in optical discharges of a moderate length is prevented by its convective nature; in the absence of the strong initial perturbations, this instability can develop, evidently, only in the optical discharge produced by weakly focused pulses.

As it follows from Eq. (8), the function $\gamma(q_x, q_y)$ reaches its maximum

$$\gamma_{\max} = \beta \varepsilon_0^{-1} - \alpha - 2k(\beta D)^{1/2} \quad (9)$$

at the point

$$q_y = 0, \; q_x = q_{\max} = (\beta k^2 / D)^{1/4}, \quad (10)$$

i.e., for the density perturbations of the kind $N_1 \sim \cos(q_{max} x)$. This perturbation presents some hybrid "diffusion–wave" periodical structure with the mixed spatial scale $\Lambda_{max} = 2\pi/q_{max} = \sqrt{2\pi \lambda L_D}$ which is the geometrical mean of the electromagnetic wavelength $\lambda = 2\pi/(k\sqrt{\varepsilon_0})$ and the characteristic diffusion length $L_D = \sqrt{D\varepsilon_0/\beta}$ (the latter is the diffusion-driven displacement of electron during the time $\varepsilon_0/\beta$). Parameter $\beta$, that determines the value of the main (positive) term in the expression for the growth rate $\gamma$, is proportional to the field amplitude derivative of the ionization rate $W$ and can be considered in the framework of our model (together with other parameters of the background state) as a slow time function. As it follows from Eq. (9), the actual instability ($\gamma_{max} > 0$) arises when this parameter becomes, in the large-scale state evolution process, larger than some threshold value. This threshold value $\beta_{th}$ and corresponding to its perturbation wave number $q_{th}$ are given by the expressions

$$\beta_{th} = \alpha \varepsilon_0 \mu^{-1} \left(1 + \sqrt{1+\mu}\right)^2, \tag{11}$$

$$q_{th} = q_{max}(\beta = \beta_{th}) = k\sqrt{\varepsilon_0(1+\sqrt{1+\mu})}, \tag{12}$$

where $\mu = \alpha/(k^2 \varepsilon_0 D)$. As we can see, the ratio of the spatial period $\Lambda_{th}$ of the unstable structure at the threshold of instability to the laser wave length in the ionized matter $\lambda$ is not a constant and is determined by the $\lambda$-depending parameter $\mu$, that is, by the ratio of the local and nonlocal (diffusion-driven) density relaxation rates ($\alpha$ and $k^2 \varepsilon_0 D$, respectively):

$$\Lambda_{th} = \lambda / \sqrt{1+\sqrt{1+\mu}}. \tag{13}$$

After passing the instability threshold, the time evolution of the spatial spectrum of the density perturbations $N_1 \sim \exp\left(\int_0^t \gamma(t') dt'\right)$ is determined by the evolution of background parameters $E_0(t)$, $N_0(t)$. Based on the known results of the numerical simulation of the laser breakdown in fused silica [6,15,16], we will consider further the comparatively simple model, in which the large-scale evolution of the background state is governed by processes of the multiphoton ionization and recombination, so that the local part of the density variation rate in Eq. (2) can be written as

$$W = W_{mpi} - \frac{N}{\tau_e}, \quad W_{mpi} = \sigma_p I^p N_a, \quad I = \frac{c\sqrt{\varepsilon_m}}{8\pi} |\mathbf{E}|^2. \tag{14}$$

Here $p$ is the number of photons required to produce one free electron; for the 800 nm radiation in the silica with the band gap of 9 eV this number is $p = 6$; $N_a = 2.1 \times 10^{22}\,\text{cm}^{-3}$ is the saturated (determined by the atom concentration) plasma density; $I$ is the laser intensity; the coefficient $\sigma_6 = 10^{-69} \text{s}^{-1} \text{cm}^{12} \text{W}^{-6}$; $\tau_e = 150$ fs is the recombination time in the fused silica [17].

It follows from Eq. (11) that the coefficients $\alpha$ and $\beta$ determining the instability growth rate are in this case

$$\alpha = \tau_e^{-1}, \quad \beta = 12 W_{mpi} / N_c, \tag{15}$$

Due to the screening action of the laser produced plasma, the dynamics of the breakdown in the focused wave beams (independently of the ionization mechanisms and the frequency bands [6,15,16,18]) is characterized by the known field self-limiting effect that is caused (at $\nu < \omega$) by the self-defocusing refraction in the field-created plasma. According to the results of above Refs. [6,15,16], under condition of interest (the 800 nm, 150 fs, ~1 μJ energy laser pulse focused to the ~1 μm waist) this effect leads to forming some quasi-steady state with the plasma density and the intensity limited in the main breakdown zone approximately at the levels $N_s \approx 0.2 N_c$, $I_s \approx 3.5 \times 10^{13}$ W/cm, respectively. As the plasma density is rather small as compared to the critical one, we set approximately, in the estimations below, $\varepsilon_0 = \varepsilon_m \approx 2$.

The least known parameter among those considered above as phenomenological ones is the ambipolar (electron-hole) diffusion coefficient $D$ defining, as well as the recombination time $\alpha^{-1} = \tau_e$, the density relaxation rate. As a rough approximation, we estimate here this coefficient based on the known general formula $D = V^2 / \nu$, replacing the effective values of the free careers thermal velocity $V$ and the collision frequency $\nu$ by the corresponding values for electrons. Following the Ref. [6,15,16], we assume the mean electron energy to be equal approximately to the energy band gap $U = 9$ eV for the fused silica to give $V \approx 2 \times 10^8$ cm/s, and following the Ref. [15] we set $\nu = 10^{14} \text{s}^{-1}$. As a result we obtain $D \approx 400$ cm$^2$s$^{-1}$, $\mu \approx 1.35$, and it is the values we take, together with the above values of the other parameters, in what follows. Of course, such simplified approach to this rather complicated issue, concerning specifically the free carrier diffusion in the fused silica at high electron temperature and strong laser field, appears as rather problematic and conjectural, but it can be evidently justified as the first step in the description of the role played by the nonlocal mechanisms in formation of the structure considered. The correct value of the equivalent diffusion coefficient $D$ in Eq. (2) (or maybe the correctness of using this phenomenological equation by itself in the laser-solid-plasma interaction) will apparently be found in the further studies. As for now, it counts in favour of the chosen above value $D$, that the associated calculations agree satisfactory with experiment.

Indeed, for the threshold ($\gamma_{max} = 0$) and steady-state (at $I_s = 3.5 \times 10^{13}$ W/cm$^2$) characteristics of the instability we obtain, respectively:

$$\beta_{th} = 6.3 \times 10^{13} \text{s}^{-1}, \quad I_{th} = 2.8 \times 10^{13} \text{ W/cm}^{-2}, \quad \Lambda_{th}/\lambda = 0.63, \quad (16)$$

and

$$\beta_s = 2.5 \times 10^{14} \text{s}^{-1}, \quad \gamma_{s\,max} = 7 \times 10^{13} \text{s}^{-1}, \quad \Lambda_{max}/\lambda = 0.45. \quad (17)$$

Both the threshold and steady-state values of the ratio $\Lambda_{max}/\lambda$ are well within the range measured in experiments [1,2]. The calculated value $\gamma_{s\,max}$ is large enough that the density perturbation can increase from the small initial level $N_1 \sim 10^{-3} N_s$ to the level $\sim N_s/2$ corresponding to the transition from the linear to the nonlinear regime in a time smaller than the pulse length. The threshold intensity $I_{th}$ lays within the same range ($\sim 10^{13}$ W/cm$^2$), where the nanograting formations were observed. It is interesting to note that this value is of the same order as those calculated in Ref. [6] ($2 \times 10^{13}$ W/cm$^2$) for the above instability of the backward scattering type.

The important parameter determining the spatial spectrum of the unstable perturbations is the characteristic width $\Delta q$ of the wave numbers region in which the instability is developed. Near the threshold, at $\Delta\beta = \beta - \beta_{th} \ll \beta_{th}$, where $\gamma_{max} = \dfrac{\Delta\beta\sqrt{1+\mu}}{\varepsilon_0(1+\sqrt{1+\mu})} \approx 2\beta_{th}\Delta I/I_{th}$, the instability takes place within a narrow band of the wave numbers in the vicinity of the point $q = q_{th}$:

$$\Delta q / q_{th} = \sqrt{\gamma_{max}/(4Dq_{th}^2)} \approx \sqrt{\Delta I/I_{th}}. \quad (18)$$

When the threshold is greatly exceeded, and in particular, at the steady-state regime with $\beta/\varepsilon_0 \gg \alpha$, $Dk^2$ the instability occupies a wide region of the wave numbers $q_1 < q < q_2$ with $q_1 \approx k\sqrt{\varepsilon_0}$, $q_2 \approx \beta/(\varepsilon_0 Dk)$. In spite of this, owing to the exponential growth of the unstable perturbations against the background steady-state, their spatial spectrum $N_1(q) = N_1^{(0)} \cdot \exp[\gamma(q)t]$ acquires a strong peak at the point $q = q_{max}$ where the growth rate is maximal. After establishing the steady-state regime, this peak is progressively increasing and sharpening in time. The time evolution of the spatial spectra of the relative density perturbations $N_1(q)/N_1^{(0)}$ is shown in Fig. 1 at the above used parameter values and the steady-state intensity $I_s = 3.5 \times 10^{13}$ W/cm$^2$. The different curves in the figure correspond to the time moments (from bottom-to-top) $t = 60, 65, 70, 75, 80$ fs; a small seed perturbation $N_1 = N_1^{(0)}$ (the same for all $q$) is given at $t = 0$. Beginning from the time $t = 70$ fs, the width $\delta q$ of the spectral peak (the half-width at half-maximum) is smaller than $0.3\, q_{max}$. If the seed density perturbation $N_i^{(0)}(x)$ is a random (delta-correlated in space) function, the space structure $N_1(x)$ corresponding to the resulted spectrum $N_1(q)$ is the quasi-sinusoidal grating with the average period $\Lambda_{max} = 0.45\lambda = 250$ nm and the correlation length is of the order of 3 periods. This separated structure evidently retains its period at the ensuing nonlinear stage of the instability, when the rate of the modulation process considered is sharply accelerated [9] as the plasma resonance condition $[(N_0 + N_1)/N_c = -\varepsilon_m]$ is approached at the points of maxima of the density perturbations.

To conclude, we have studied the linear stage of the small-scale ionization-induced instability that develops within the volume of the transparent dielectric (fused silica) during the process of its ionization by the fs laser pulse. This instability results in the 1D spatial modulation of the plasma density in the direction of the laser electric field and evidently can be considered as the initial cause (alternative to one previously considered in Refs. [2,3,13,14]) of the nanograting formation produced in silica by the series of the laser pulses. Though the suggested model allows to explain the observed periodicity of the formed plasma structure (as some hybrid nonlinear diffusion-wave formation), it relates only to the first (concerning the plasma behavior) part of the grating formation problem. More complete study should probably include also the analysis of the cumulative thermochemical processes in dielectric that can, in its turn, affect the plasma parameters governing the instability considered (diffusion coefficient, times of recombination, effective collision frequency). Probably, only with these processes taken into account, it occurs to be possible to explain the observed (but still not understood) dependences of the grating period on the parameters of the irradiation regime.

This work was supported by the Government of the Russian Federation (Agreement No. 14.B25.31.0008) and the Russian Foundation for Basic Research (Grant No. 14-02-00847).

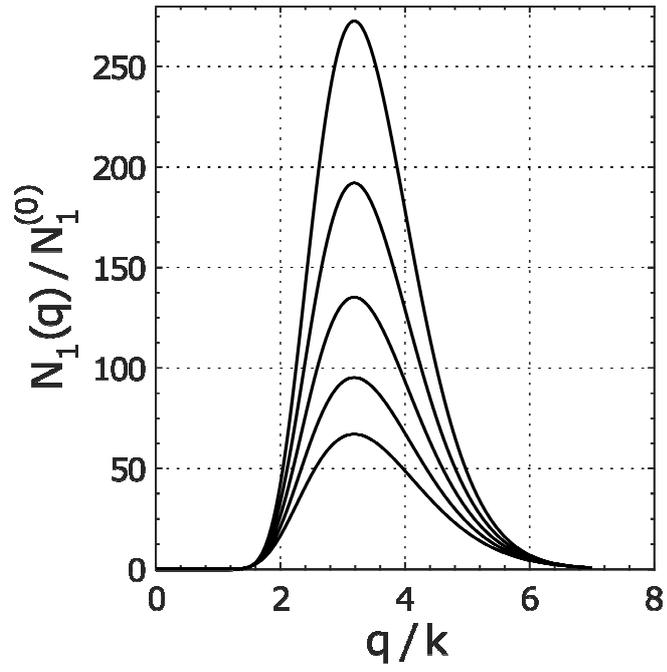

Fig. 1. Spatial spectra of the plasma density perturbations at different time instants changing with the step 5 fs from 60 fs (the bottom curve) to 80 fs (the top one); the seed perturbation $N_1 = N_1^{(0)}$ is given as a constant at the moment $t = 0$. The laser intensity is $3.5 \times 10^{13}\,\text{W/cm}^2$, the wavelength is 800 nm.